\documentclass[twocolumn,showpacs,preprintnumbers,amsmath,amssymb]{revtex4}
\usepackage{epsf,amsmath,amssymb,verbatim,color,multirow,pifont}
\usepackage{graphicx}

\begin{document}

\title{Jamming transition in traffic flow under the priority queuing protocol}
\author{K.~\surname{Kim}}
\author{B.~\surname{Kahng}}
\author{D.~\surname{Kim}}
\affiliation{Department of Physics and Astronomy, Seoul National
University, Seoul 151-747}
\date{\today}

\begin{abstract}
Packet traffic in complex networks undergoes the jamming
transition from free-flow to congested state as the number of
packets in the system increases. Here we study such jamming
transition when queues are operated by the priority queuing
protocol and packets are guided by the dynamic routing protocol.
We introduce a minimal model in which there are two types of
packets distinguished by whether priority is assigned. Based on
numerical simulations, we show that traffic is improved in the
congested region under the priority queuing protocol, and it is
worsened in the free-flow region. Also, we find that at the
transition point, the waiting-time distribution follows a power
law, and the power spectrum of traffic exhibits a crossover
between two $1/f^{\alpha}$ behaviors with exponent $\alpha \approx
1$ and $1 < \alpha < 2$ in low and high frequency regime,
respectively. This crossover is originated from a characteristic
waiting time of packets in the queue.
\end{abstract}

\maketitle

Information packet transport via the Internet is an important
problem in complex systems from the perspectives of both theory
and application. Data packets created at certain nodes in the
Internet travel to their destinations under transmission control
protocols. During the journey, packets interact with other packets
as they share a common line or buffer in the network. Accordingly,
several types of collective behaviors can emerge in the form of
self-similar traffic~\cite{fekete, willinger} or
chaos~\cite{veres}. To enhance transport efficiency, one would
like to design an appropriate protocol to transport as many
packets as possible with the lowest cost. With this goal in mind,
several packet transport models have been introduced.

When a packet is sent from one node to another in a network, it is
usually routed along the shortest path; such a path is undoubtedly
the best route when the number of packets in the network is
relatively small. However, when many packets are floating around
in the network, traffic congestion can occur. This problem can be
especially serious in scale-free (SF) networks~\cite{ba}, since
hubs are the bottlenecks of traffic flow. To resolve this
congestion, many routing protocols have been proposed, including
the hub avoidance protocol~\cite{sre,yan} and the optimal routing
protocol~\cite{danila}. These are static routing protocols, so
that a path from one node to another is fixed regardless of the
traffic level in the system at any given time. In contrast, there
is a dynamic routing protocol that guides packets to alternative
paths depending on the traffic on the path to each
target~\cite{ech}.

When a packet travels through a node (router), it is temporarily
stored in the buffer (queue) at the node. There can be many
queuing protocols that control the order of packet transmission in
the queue. The most common one is the `first-in-first-out' (FIFO)
protocol. Alternatively, the `last-in-first-out' (LIFO) protocol
can be used~\cite{Tad1,Tad2,Tad3}. The priority queue is a rather
different protocol~\cite{priority}. Each packet is assigned a
priority upon its birth. A packet with the highest priority is
treated first in the queue, irrespective of its order of arrival.
The diffusion process~\cite{maragakis} under the priority queuing
protocol in complex networks have been studied previously.

In this paper, we study the jamming transition of packet transport
on SF networks using the priority queuing protocol and the dynamic
routing protocol by adapting Dijkstra's algorithm~\cite{dijk}. At
each time step, every node creates a packet with probability $q$
whose destination is chosen randomly. These packets are assigned
to be either with or without priority. The fraction of the packets
that are priority-assigned is $f$. In the queue, packets with
priority are delivered first, followed by packets without
priority. Similar types of packets in the queue are treated
following the FIFO protocol. Packets with priority may be regarded
as paid packets when downloaded from a certain web site. Note that
when $f=0$ or $f=1$, the priority queuing protocol reduces to the
standard FIFO protocol. We present a phase diagram for the
free-flow and congested phases in the parameter space ($q$,$f$),
and we show that the priority queuing protocol is efficient when
the system is congested.

We simulate packet transport under the dynamic rules below on
undirected binary scale-free networks generated by the static
model~\cite{load}. For the network, the total number of nodes is
$N=1,000$, the average degree of a node is $\langle k \rangle
\simeq 4$, and the degree exponent is $\gamma=2.5$. Each type of
packet travels along the path that minimizes the quantity
\begin{equation}
L_{s,d}(t) = \ell_{x} + h \sum_{i\in x} Q_{i}(t), \label{hh}
\end{equation}
where $\ell_{x}$ is the hopping distance along a path $x$ between
nodes $s$, $d$, $Q_{i}(t)$ is the queue length at node $i$ on the
path $x$, and $h$ is a traffic-control parameter~\cite{ech}. For
packets with (without) priority, $Q_{i}(t)$ is regarded as the
number of priority-assigned (both types of) packets that have
accumulated in the queue at node $i$. Hence, the path minimizing
$L_{s,d}$ can be the best choice to route the packet at time $t$,
since the packet can circumvent congested nodes along its way.
This path is determined using Dijkstra's algorithm in the
simulation. Note that, when $h = 0$, this routing protocol reduces
to the shortest path routing protocol. We choose other control
factors as follows: The queue size is unlimited, and the process
rate in each queue is one packet per time step. Thus, if more than
one packet arrives at a node per unit time, then the queue length
increases. The system is updated in parallel, meaning that all
packets move simultaneously with the queue length information of
the previous time step. Our simulations are performed for up to
$10^4$ time steps.

To characterize the jamming transition, we define an order
parameter for each type of packet as the delivery fraction
$D_{\alpha}$, where $\alpha ={\rm p}$ for packets with priority,
$\alpha={\rm n}$ for packets without priority, and $\alpha={\rm
tot}$ for all packets combined. $D_{\alpha}$ is defined as
\begin{equation}
D_{\alpha}=\lim_{t \rightarrow
\infty}\frac{1}{N_{\alpha}(t-t_{0})}\int_{t_0}^{t}\lambda_{\alpha}(t^{\prime})dt^{\prime},
\end{equation}
where $\lambda_{\alpha}(t)$ is the number of packets of type
$\alpha$ arriving at their destinations at time $t$ and
$N_{\alpha}$ is the number of packets of type $\alpha$ generated
in a unit time. Here, $N_{\alpha}$ is $N_{\alpha} = Nqf$,
$Nq(1-f)$, and $Nq$ for $\alpha =$ {\rm p}, {\rm n}, and {\rm
tot}, respectively. If the traffic of packets of type $\alpha$ is
in the free-flow state, then $D_{\alpha}$ is close to 1, since all
packets of type $\alpha$ arrive successfully at their targets.
However, if the traffic is in a congested state, then
$0<D_{\alpha}<1$. When $D_{\alpha}=0$, the traffic is in the
completely congested state.

\begin{figure}[t]
\includegraphics[width=5cm, angle=-90]{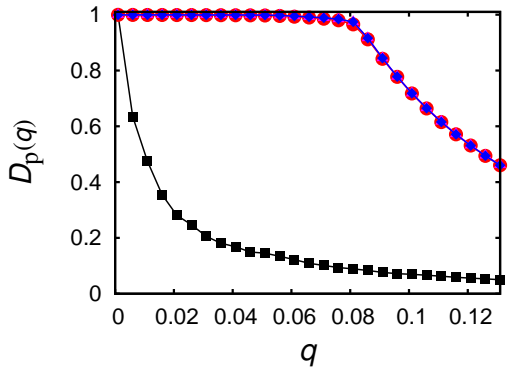}
\caption{(Color online) The delivery fraction $D$ for packets with
priority versus the packet generation rate $q$ for various $h$.
The traffic-control parameter $h$; $h=0.0$ ($\square$), $0.1$
($\bigcirc$), $0.5$ ($\bigtriangleup$) and $1.0$ ($\diamondsuit$).
The priority fraction $f$ is chosen as $0.9$. }\label{traffic}
\end{figure}

In general, the jamming transition point for packets with priority
does not coincide with that for packets without priority, denoted by
$q_{\rm p}$ and $q_{\rm n}$, respectively.
Obviously $q_{\rm p}> q_{\rm n}$. When $f=0$ and $f=1$, all
packets are of the same type and the queuing process is governed
by the FIFO protocol. In this case, the delivery fraction and the
jamming transition point are denoted by $D_0$ and $q_0$,
respectively.

First, we consider the effect of the traffic-control parameter
$h$. We observe that the traffic is dramatically improved when $h
> 0$ compared with the case when $h=0$, as shown in
Fig.~\ref{traffic}. However, the traffic seems to be unaffected by
changes in $h$ when $h > 0$. This result is reasonable because the sum
of accumulated packets along the path is far larger than the
topological length $\ell_x$ in the congested state. Hence, we will
confine our interest to the case when $h=1$ from here on.

\begin{figure}[t]
\includegraphics[width=5cm, angle=-90]{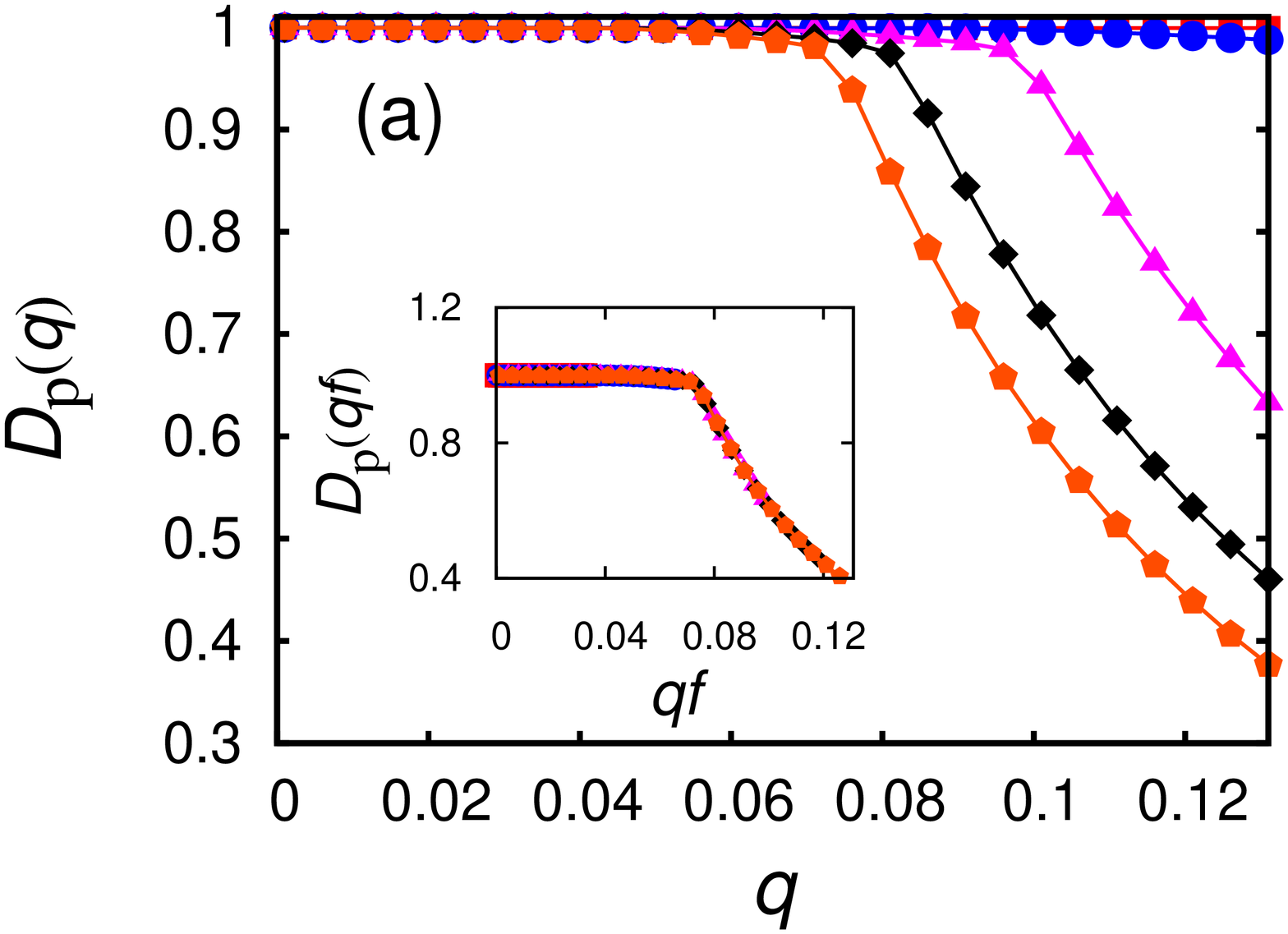}
\includegraphics[width=5cm, angle=-90]{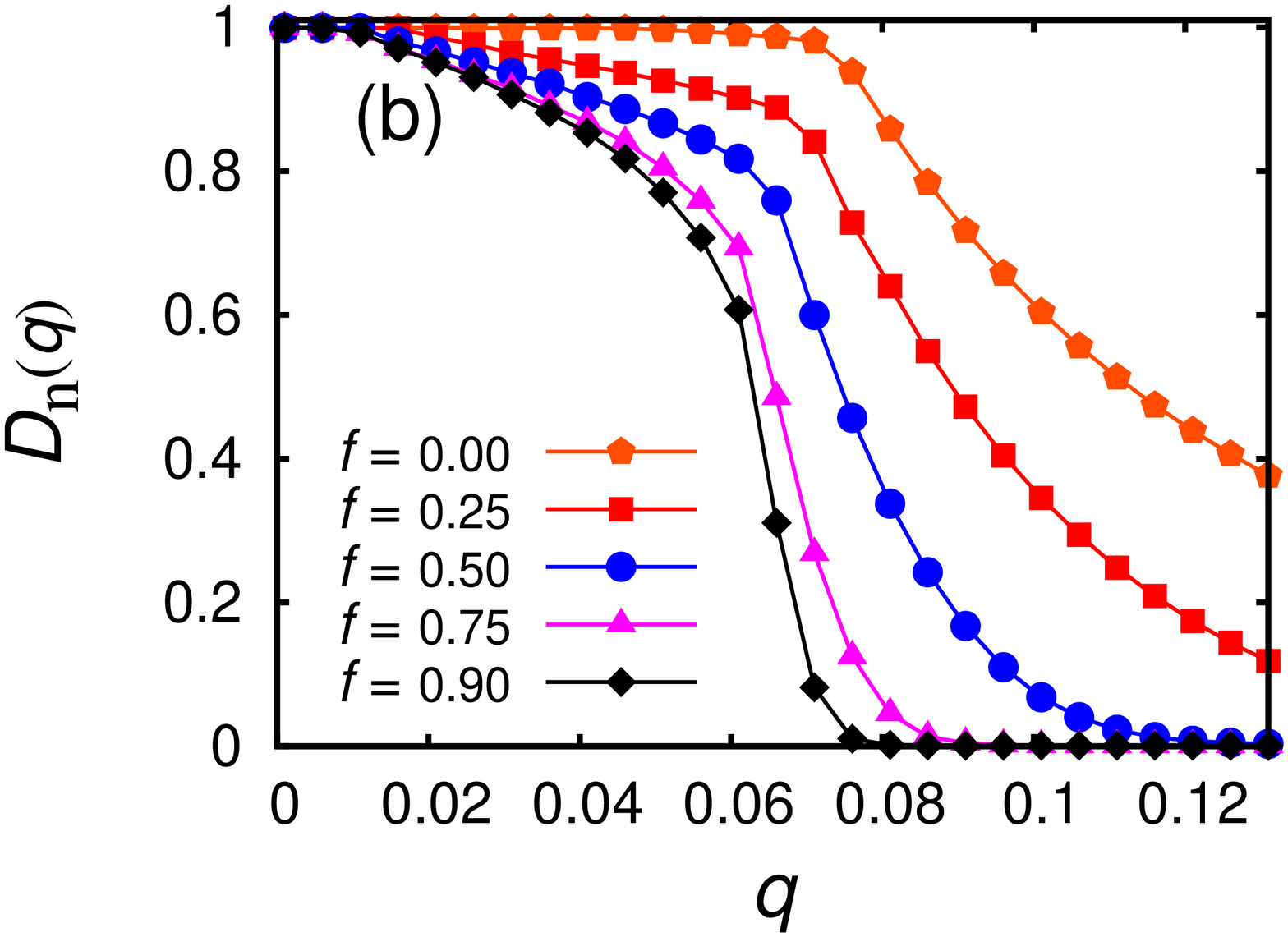}
\includegraphics[width=5cm, angle=-90]{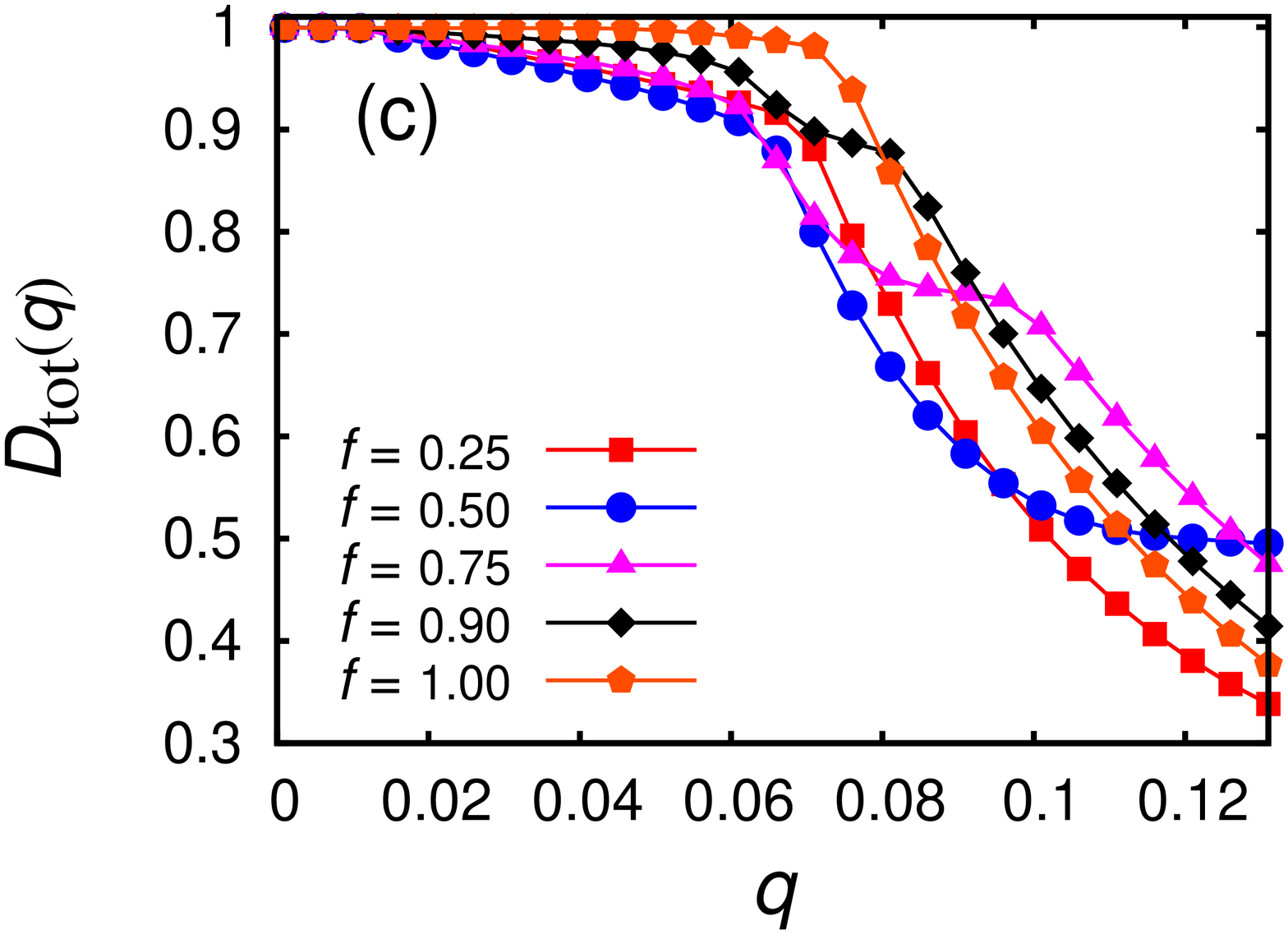}
\caption{(Color online) The delivery fraction $D$ for packets with
(a) and without priority (b), and all packets (c) versus the
packet generation rate $q$ for various fractions $f$ of the
priority assignment. The inset at (a) shows $D_{\rm p}$ collapses
a unique scaling function.} \label{deliver}
\end{figure}

\begin{figure}[t!]
\includegraphics[width=5cm, angle=-90]{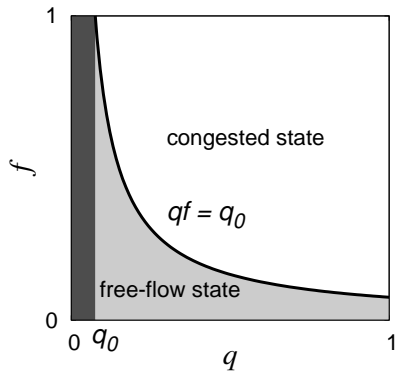}
\caption{Phase diagram for packet traffic. The dark grey region
represents the free-flow state when the FIFO protocol is used. The
light grey region represents the additional free-flow state for
the priority-assigned packets under the priority queuing protocol.
}\label{phase}
\end{figure}

We observe the behavior of the delivery fraction. We consider
$D_{\rm p}$ as a function of $q$ for various $f$ in
Fig.~\ref{deliver}(a). The preliminary result of this was reported
in~\cite{kim}. For $f=0$ and $f=1$, we obtain $q_0 \simeq 0.07$,
which can change depending on the system size $N$. When $0 < f <
1$, the jamming transition point $q_{\rm p}$ for priority-assigned
packets is larger than $q_0$. Since packets without priority do
not hamper the traffic of packets with priority, one can obtain
the relation
\begin{equation}
 q_{\rm p} f =q_0.
\end{equation}
On the other hand, when $q > q_{\rm p}$, a fraction of packets with priority
cannot reach their targets during a
given time interval. Hence, $D_{\rm p} < 1$ in such a case.

Second, we examine the behavior of the jamming transition for
packets without priority. Since packets with priority delay the
traffic of packets without priority under the priority queuing
protocol, the jamming transition for packets without priority
occurs at $q_{\rm n}$ smaller than $q_0$. As $q_{\rm p} > q_{\rm
n}$ for a given $f$, $D_{\rm p}=1$ and $0 < D_{\rm n} < 1$ for
$q_{\rm n} < q < q_{\rm p}$. Fig.~\ref{deliver}(b) shows the
behavior of $D_{\rm n}$ as a function of $q$ for various $f$. For
$q > q_{\rm p}$, the traffic for packets without priority is
completely congested, i.e., $D_{\rm n}\simeq 0$.

Next, we combine the above two cases and consider the jamming
transition for all packets irrespective of priority assignment.
$D_{\rm tot}$ is the order parameter for all packets. The behavior
of $D_{\rm tot}$ is shown in Fig.~\ref{deliver}(c). $D_{\rm tot}$
satisfies the relationship
\begin{equation}
D_{\rm tot} = fD_{\rm p} + (1-f)D_{\rm n},
\end{equation}
where $D_{\rm p}$ and $D_{\rm n}$ can change depending on $q$. We
summarize the delivery fraction for all packets as
\begin{equation}\label{dt}
D_{\rm tot}=\left\{\begin{array}{ll} 1 &~~~{\rm for}~~q < q_{\rm n}, \\
f + (1-f)D_{\rm n} &~~~{\rm for}~~q_{\rm n} < q < q_{\rm p}, \\
fD_{\rm p} &~~~{\rm for}~~q > q_{\rm p}. \end{array} \right.
\end{equation}

It is interesting to note that the delivery fraction $D_{\rm tot}$
under the priority queuing protocol can exceed the $D_0$ obtained
from the simple FIFO protocol. This phenomenon can occur in the
region of $q > q_{\rm p}$, as shown in Fig.~\ref{deliver}(c) (for
example, $q > 0.12$ for $f = 0.5$, and $q > 0.096$ for $f =
0.75$). The unexpected improvement in overall transport efficiency
is due to the priority queuing protocol, since it enables us to
control the density of packets with priority; these packets remain
deliverable packets in the congested state. For the other cases,
$D_{\rm tot} < D_0$, implying that the overall traffic under the
priority queuing protocol is worse in the free-flow state, despite
the improvements observed when the system is in the congested
state.

\begin{figure}
\includegraphics[width=5cm, angle=-90]{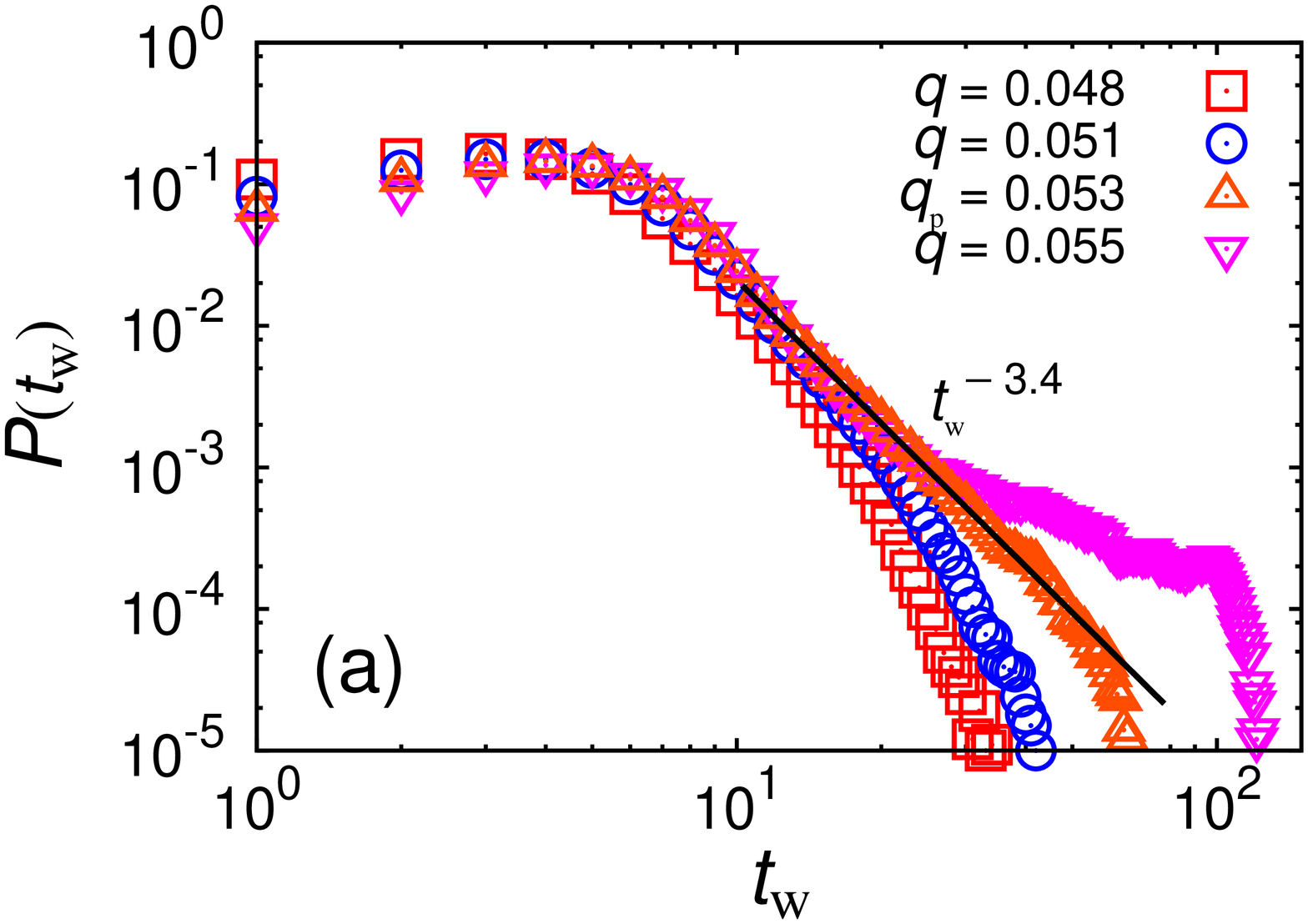}
\includegraphics[width=5cm, angle=-90]{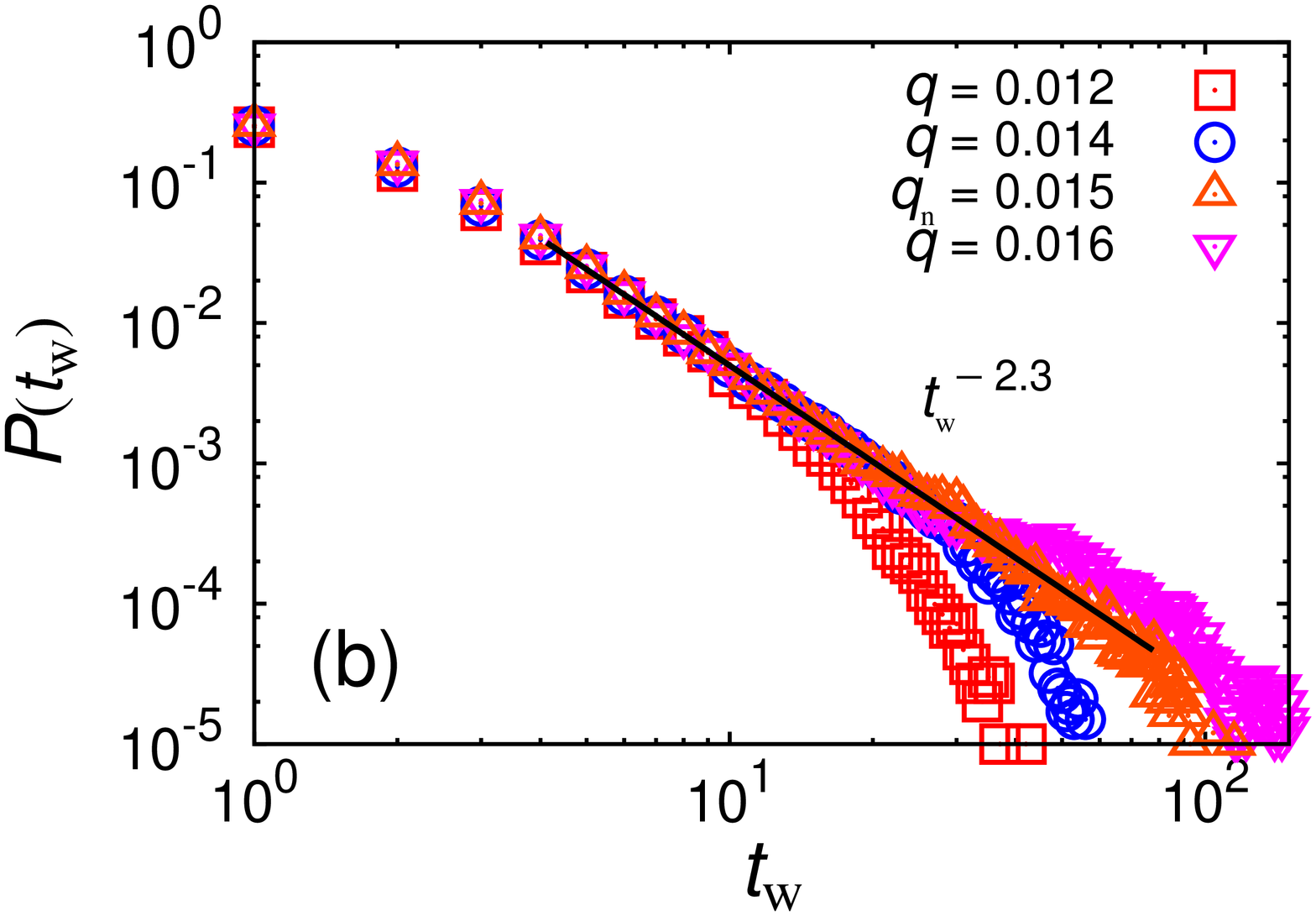}
\caption{(Color online) The waiting-time distribution in the
free-flow region for packets with (a) and without (b) priority for
$f = 0.9$ and $f = 0.25$ respectively. The power-law behaviors are
obtained at $q_{\rm p}=0.053$ and $q_{\rm n}=0.015$, which are
regarded as the jamming transition points. The solid lines are
guidance with slopes $-3.4$ (a) and $-2.3$ (b),
respectively.}\label{waiting}
\end{figure}

Fig.~\ref{phase} is a phase diagram of the traffic of packets in the space of
$(q,f)$. Under the FIFO protocol only, the phase space is divided into two parts
by the line $q=q_0$, with the free-flow state appearing in the region $q < q_0$ and
the congested state appearing in the region $q > q_0$. However, when
the priority queuing protocol is used, the free-flow region can be
extended into the region $q < (q_0/f)$.

\begin{figure}
\centering
\includegraphics[width=5cm, angle=-90]{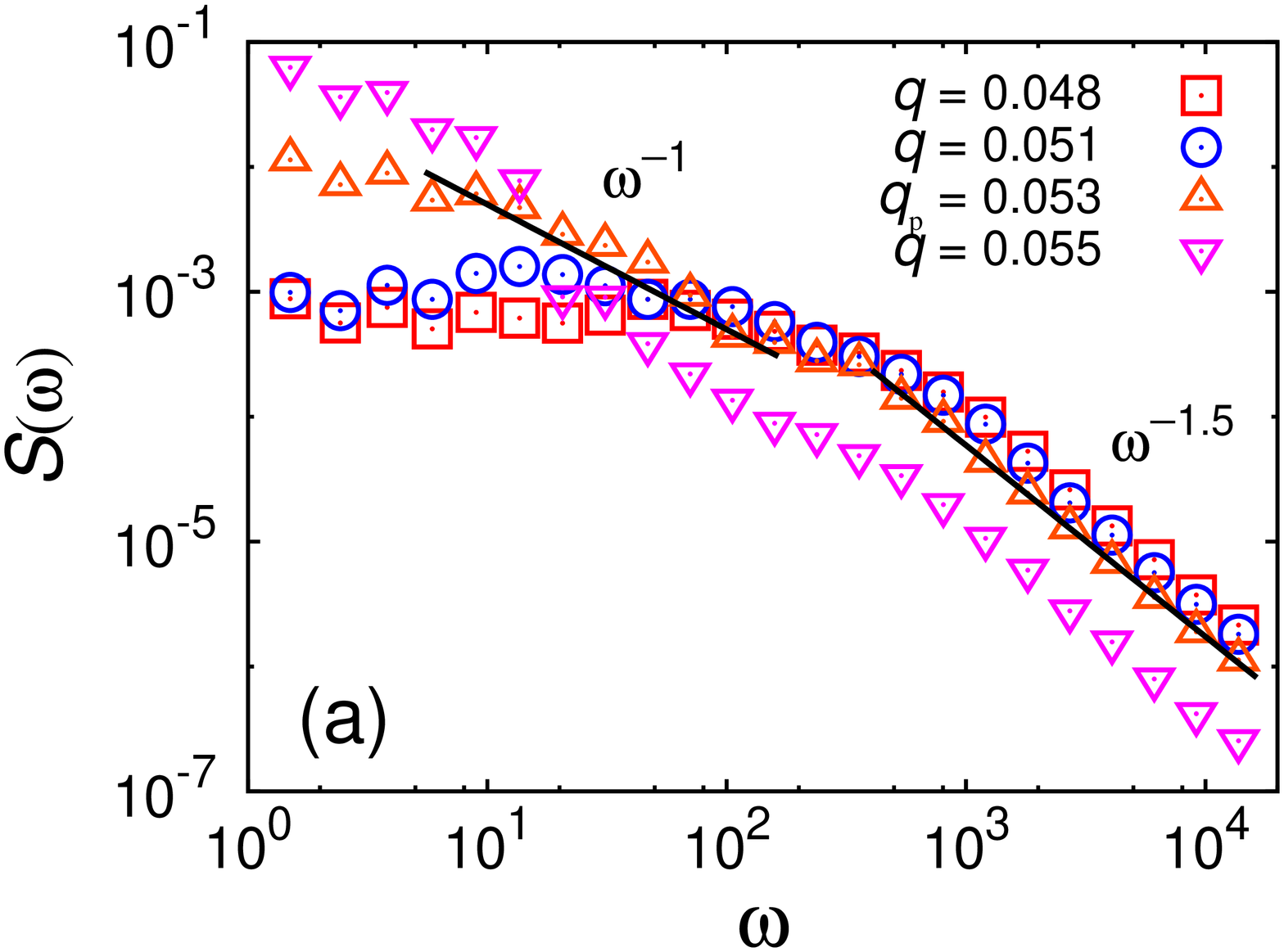}
\includegraphics[width=5cm, angle=-90]{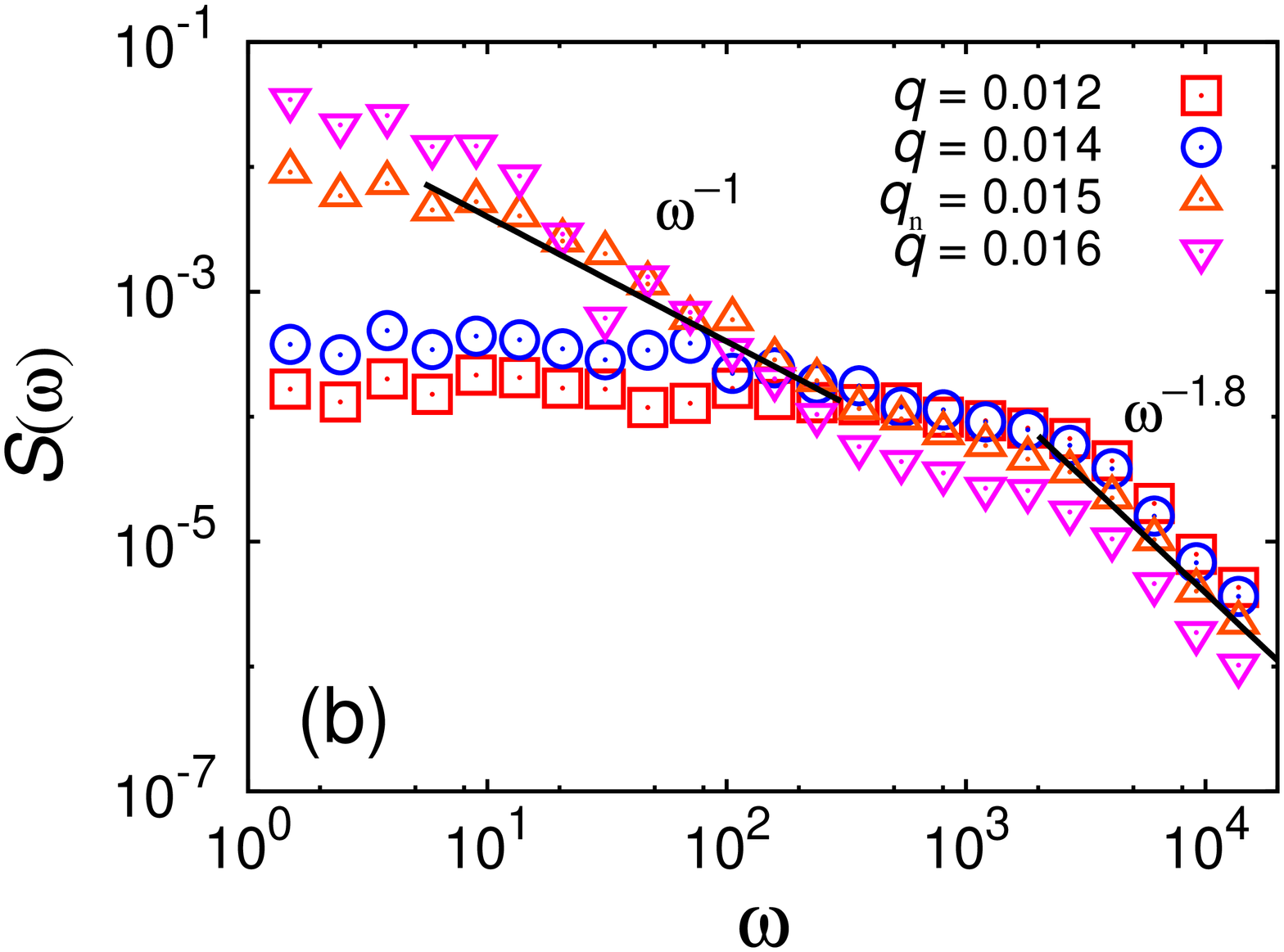}
\caption{(Color online) The power spectrum of traffic in the
system for packets with (a) and without (b) priority,
respectively. We use $f = 0.9$ (a) and $f = 0.25$ (b). For $q_{\rm
p}=0.053$ (a) and $q_{\rm n}=0.015$ (b), the power spectrum
exhibits a crossover between two power-law behaviors. Solid lines
are guidelines with slopes $-1$ in small $\omega$ regime and
$-1.5$ in large $\omega$ regime for (a) and $-1$ in small $\omega$
regime and $-1.8$ in large $\omega$ regime for (b).}\label{power}
\end{figure}

We measure the waiting-time distributions $P(t_{\rm w})$ for
packets with and without priority~\cite{Tad1,Tad2}. The waiting
time $t_{\rm w}$ is defined as the time spent in the queues on the
way to the target, excluding the transit time. As shown in
Fig.~\ref{waiting}, the waiting-time distribution follows a power
law $P(t_{\rm w})\sim t_{\rm w}^{-\delta}$ near the transition
points $q_{\rm p}$ and $q_{\rm n}$, where $\delta\approx 3.4(1)$
for packets with priority and $\delta \approx 2.3(1)$ for packets
without priority. In the free-flow regions $q < q_{\rm p}$ and $q
< q_{\rm n}$, the waiting-time distribution behaves as $P(t_{\rm
w}) \sim e^{-t_{\rm w}/\tau}$, where $\tau$ is the mean waiting
time and depends on the density of packets in the network. The
maximum waiting time of packets in the free-flow region is
estimated to be $t_m\approx 70$ for a system of size $N=10^3$.
This means that packets can reach their targets within $t_m$ at
the most.

The power spectrum $S(\omega)$ of the traffic is defined as
\begin{equation}
S(\omega)=\frac{|g(\omega)|^2}{\sum_{\omega=0}^{T/2}|g(\omega)|^2},
\end{equation}
where $g(\omega)=\sum_{t=0}^{T-1} F(t) e^{-{\rm
i}\frac{2\pi\omega}{T}t}$ and $F(t)$ is the number of packets with
(without) priority in the system at time $t$. The power spectrum
of the traffic in the system is measured in Fig.~\ref{power}. The
behavior of the power spectrum $S(\omega)$ depends on the packet
generation rate $q$ as well as the packet type. When the packet
generation rate $q$ is near the jamming transition point, the
power spectrum exhibits a crossover between two power-law
behaviors $S(\omega) \sim \omega^{-\eta}$ with $\eta\approx 1$ and
$\eta\approx 1.5(1)$ for packets with priority and $\eta\approx 1$
and $\eta\approx 1.8(1)$ for packets without priority. Such
behaviors are different from what is observed in model systems and
empirical data~\cite{Tad2,takayasu}. This difference is probably
due to the short-tailed behavior of $P(t_{\rm w})$ compared with
that for~\cite{Tad2, takayasu}. This difference of $P(t_{\rm w})$
is caused by the difference in routing and queuing protocol. The
crossover behavior occurs roughly at $\omega_c\approx T/2\pi
t_m\approx 10^2$, where $T = 5\times 10^4$ is the total simulation
time step. This value corresponds to the maximum waiting time in
the system, $t_{\rm m}$, roughly estimated in Fig.~\ref{waiting}
to be $60\sim 70$ for a system of size $N=10^3$. The $1/f$-type
power spectral density suggests that there exists a long time
correlation in the transport of both types of packets.

We also perform the same simulations on the Erd\H{o}s and Ren\'yi (ER)
network to ascertain whether our result is affected by network structure.
With the exception of the increment of the transition point, the generic features
of the simulation results remain unchanged using the ER network, indicating that
our results are independent of network structure.

In summary, we studied the packet transport problem on SF networks
under the priority queuing protocol and the dynamic routing
protocol. We showed that total traffic can be improved in the
congested state by introducing the priority queuing protocol,
although the overall traffic is worse in the free-flow
state and the jamming transition point is reduced. The jamming
transition points for packets with and without priority are
different. Near each jamming transition point, the waiting-time
distribution follows a power law, and the power spectrum exhibits
a crossover between two power-law behaviors. We obtain $1/f$-type power
spectra in the small $\omega$ regime for both types of
packets.

This work was supported by KOSEF grant Acceleration Research (CNRC) (No.R17-
2007-073-01001-0).

\end{document}